# Thermal Dissipation and Variability in Electrical Breakdown of Carbon Nanotube Devices

Albert Liao[1,2], Rouholla Alizadegan[3], Zhun-Yong Ong[1,4], Sumit Dutta[1,2], Feng Xiong[1,2], K. Jimmy Hsia[1,3] and Eric Pop[1,2,5,*]

[1]*Micro and Nanotechnology Lab, Univ. Illinois at Urbana-Champaign, Urbana IL 61801*
[2]*Dept. of Electrical & Computer Eng., Univ. Illinois at Urbana-Champaign, Urbana IL 61801*
[3]*Dept. of Mechanical Science & Eng., Univ. Illinois at Urbana-Champaign, Urbana IL 61801*
[4]*Dept. of Physics, Univ. Illinois at Urbana-Champaign, Urbana IL 61801*
[5]*Beckman Institute, Univ. Illinois at Urbana-Champaign, Urbana IL 61801*

We study high-field electrical breakdown and heat dissipation from carbon nanotube (CNT) devices on SiO$_2$ substrates. The thermal "footprint" of a CNT caused by van der Waals interactions with the substrate is revealed through molecular dynamics (MD) simulations. Experiments and modeling find the CNT-substrate thermal coupling scales proportionally to CNT diameter and inversely with SiO$_2$ surface roughness ($\sim d/\Delta$). Comparison of diffuse mismatch modeling (DMM) and data reveals the upper limit of thermal coupling ~0.4 WK$^{-1}$m$^{-1}$ per unit length at room temperature, and ~0.7 WK$^{-1}$m$^{-1}$ at 600 $^{\circ}$C for the largest diameter (3-4 nm) CNTs. We also find semiconducting CNTs can break down prematurely, and display more breakdown variability due to dynamic shifts in threshold voltage, which metallic CNTs are immune to; this poses a fundamental challenge for selective electrical breakdowns in CNT electronics.

## I. Introduction

Carbon nanotubes (CNTs) have excellent intrinsic electrical and thermal properties, and thus are being considered potential candidates for nanoscale circuits,[1] heat sinks[2] or thermal composites.[3] However, their physical properties depend on temperature, and thus are directly affected by power dissipation during electrical operation.[4-6] Joule heating in CNTs goes beyond degrading electrical performance, posing reliability concerns as in other electronics. Electrical Joule breakdown has also been used to remove metallic CNTs in integrated circuits;[7-9] however the technique is not precise, owing to the lack of fine control over CNT heat dissipation. It is presently understood that the thermal boundary conductance (TBC) at CNT interfaces with the environment, substrate, or contacts plays the limiting role in thermal dissipation.[10-12] In addition, the interaction of CNTs with the environment may also change their effective thermal conductivity.[13, 14] However, little is currently known about the details of the thermal interaction between CNTs and common dielectrics, including the roles of dielectric surface roughness or of CNT diameter and chirality (e.g. metallic vs. semiconducting).

In this study, we examine electrical breakdown and thermal dissipation of CNT devices with the most common interface used in integrated circuit experiments, that of $SiO_2$ as shown in Fig. 1(a). We employ electrical breakdown thermometry[11, 15] to extract the TBC between CNTs and $SiO_2$ for metallic (m-CNT) and semiconducting nanotubes (s-CNT) of diameters $1 < d < 4$ nm. We find the TBC per unit length scales proportionally with CNT diameter, confirming recent simulation work.[16] We also find that m-CNTs appear to have better and more consistent thermal coupling with $SiO_2$ than s-CNTs, indicating a fundamental challenge for complete m-CNT removal in circuits via electrical breakdowns. We compare our results to both a diffuse mismatch model (DMM) and to molecular dynamics (MD) simulations. The latter reveal the role played by the thermal "footprint" of a deformable CNT on such dielectric substrates. Finally, we uncover the significant role of variability in threshold voltage (for s-CNTs) and of $SiO_2$ surface roughness (for both m- and s-CNTs) in heat dissipation and electrical breakdown.

## II. Experiments and Data Extraction

We fabricated and conducted experiments on carbon nanotube devices in the same back-gated configuration as our previous work, using semi-circular electrodes for better CNT length control[17-19] (here, $2 \leq L \leq 5.6$ µm) as shown in Figs. 1 and 2. The $SiO_2$ is thermally grown dry oxide, approximately 90 nm thick. We focused on nanotubes that showed high-bias current near ~25 µA[20] and had diameters $d < 4$ nm as measured by atomic force microscopy (AFM), to ensure devices were single-walled. In addition, only electrical breakdowns with a single, clean drop to zero current were selected, which are typical of single-wall single-connection devices, as shown in Fig. 1(b); by contrast, multi-wall CNTs and CNT bundles exhibit higher currents and break down with multiple current steps.[21] Joule heating was achieved by increasing the source-drain voltage ($V_{SD} > 0$) while maintaining a negative gate bias ($V_{GD} \approx -15$ V). In semiconducting CNTs this leads to hole-only conduction,[17] deliberately avoiding ambipolar behavior[22] which would complicate the analysis. Metallic CNTs show no gate voltage dependence in room temperature, ambient conditions. Increasing $V_{SD}$ leads to increasing the power input, which causes the CNT temperature to rise through Joule heating and leads to physical breakdown. We note that in this work the drain is always grounded and the source is the positive terminal, referring to the source of carriers and current flow.

The breakdown voltage, $V_{SD} = V_{BD}$ is the voltage at which the drain current ($I_D$) irreversibly drops to zero, as shown in Fig. 1(b). Typical broken devices under AFM imaging are shown in Figs. 1(c) and 2(a). The power dissipated within the CNT at breakdown is $P_{BD} = I_D(V_{BD} - I_D R_C)$. The combined resistance of the source and drain contacts, $R_C$, is estimated from the inverse slope of the low-bias $I_D$-$V_{SD}$ plot,[11, 23] $R_C \approx (dI_D/dV_{SD})^{-1}$, which includes the quantum contact resistance ($R_0 = 6.5$ kΩ). The experiments in this study were performed in air where nanotubes are known to break from self-heating and oxidation at a relatively well-known temperature,[24] $T_{BD} \approx 600°C$. By comparison, device breakdowns performed in ~$10^{-5}$ Torr vacuum showed CNTs of similar lengths and diameters breaking at approximately three times higher power and thus at temperatures near ~1800 °C, as in Fig. 1(b). This suggests that CNT device breakdowns in vacuum occur by a mechanism other than oxidation, e.g. at nanotube defects[25] or by failure of the underlying $SiO_2$ (which begins to melt at ~1700 °C, consistent with the higher CNT power at breakdown). The latter is supported by the observation of damage or reflow of the $SiO_2$ substrate in some samples, as seen in Fig. 1(c), which is never seen for breakdowns in air.

We now return to discuss the temperature profile of CNTs during Joule heating, and restrict ourselves to in-air breakdowns for the rest of the manuscript. Figure 2(a) displays the breakdown location ($L_{BD}$) along a CNT, as extracted from scanning electron microscope (SEM) imaging.

Figure 2(b) shows a histogram of the normalized breakdown locations for ~40 CNTs in this study, distinguishing between m-CNT and s-CNTs. The majority of m-CNTs break at their hottest point near the middle while most s-CNTs break closer to the grounded drain, where the field is higher and the carrier density is lower. Both of these observations are indicative of diffusive heat[11] and charge[26] transport, and of relatively negligible contact resistance. At high field the electron or hole scattering mean free path (MFP) with optical phonons (OP) approaches the minimum value $\lambda_{OP,ems}$ ~15$D$ where $d$ is the diameter in nm.[17, 23] This MFP is significantly shorter than the CNT lengths used in this work (several microns).

To understand the temperature profiles of m-CNTs and s-CNTs, and to extract the interfacial thermal conductance per unit length ($g$) between CNT and $SiO_2$ from the breakdown data, we solve the heat diffusion equation along the CNT.[11] The heat generation per unit length can be captured both as uniform (for m-CNTs) and asymmetric (for s-CNTs), by expressing it as:

$$p(x) = p_0 \left( C_1 + \frac{C_2 x}{L} \right) \quad (1)$$

where $-L/2 \leq x \leq L/2$, $L$ is the length of the CNT, $C_1$ and $C_2$ are unitless parameters and $p_0$ is a constant term. We note that to a good approximation the heat generation in CNTs is independent of temperature, as the optical phonon emission length (the strongest inelastic scattering mechanism responsible for Joule heating) has very weak temperature dependence.[11, 17]

For m-CNTs the heat generation is uniform due to constant electric field and charge density (barring significant and asymmetric contact resistance[27, 28]) we simply set $C_1 = 1$ and $C_2 = 0$. This implies $p_0 = P_{BD}/L$ at breakdown in m-CNTs. For s-CNTs, a linear heat generation profile captures the asymmetry caused by non-uniform electric field and charge density.[29] The general expression for the temperature along the CNT at breakdown is:

$$T(x) = T_0 + \frac{P_{BD}}{g_{tot} L} \left[ C_1 + \frac{C_2 x}{L} - \frac{C_1 \cosh\left(\frac{x}{L_H}\right)}{\cosh\left(\frac{L}{2L_H}\right)} - \frac{C_2 \sinh\left(\frac{x}{L_H}\right)}{2\sinh\left(\frac{L}{2L_H}\right)} \right] \quad (2)$$

where $L_H = (kA/g)^{1/2}$ is the thermal healing length (of the order ~0.2 μm),[11, 30] $k$ is the thermal conductivity of the CNT,[26] $g_{tot}$ is the thermal conductance per unit length from CNT to ambient (see Section III below), and $A = \pi a d$ is the cross-sectional area assuming a CNT wall thickness $a = 0.34$ nm.

The typical "inverted U" shape of the temperature profile under uniform heat generation in m-CNTs is shown in Fig. 2(c) with $C_2 = 0$. This has previously been observed experimentally in nanotubes under high bias operation, both by scanning thermal microscopy (SThM)[31] and by coating the CNTs with a phase-change material which changes volume as it heats up.[30]

On the other hand, s-CNTs have non-uniform electric field and charge density along their length, leading to off-center heat dissipation.[29] This is captured by changing the value of the parameter $C_2 > 0$ above, as shown in Fig. 2(c). We take this simple approach because uncertainties in threshold voltage, contact resistance, and contributions made by infrequent defects make it difficult to provide a more exact solution of the temperature profile in every s-CNT measured. (by contrast, m-CNTs are immune to threshold voltage variations). More specifically, in our analysis below we choose $C_1 = 1$ and $C_2 = 0.65$ for s-CNTs, such that the hot spot location corresponds to $L_{BD}/L \sim 0.7$ as noted in the breakdown histogram, Fig. 2(b).

## III. Diffuse Mismatch Model

To understand the dependence of thermal coupling $g$ on CNT diameter and substrate properties we use a diffuse mismatch model (DMM)[32] in a similar manner previously applied to multi-wall carbon nanotubes[33] and graphene.[34] The DMM is used to establish an upper bound for heat transport across an interface, as limited by the phonon density of states (PDOS). This approach also presents an advantage of speed and flexibility over full MD methods.[16] The model calculates the transmission probability, $\tau$, for heat transfer across an interface while assuming all phonons scatter diffusely at the interface. By equating the phonon energy flux from the CNT to the SiO$_2$ with that from the SiO$_2$ to the CNT and using a detailed balance argument for all frequencies,[34,35] $\tau$ is given as:

$$\tau = \frac{\frac{1}{4}\int \omega N_{OX} f_{BE,OX} D_{OX} v_{ox} d\omega}{\frac{\int \omega N_{CNT} f_{BE,CNT} D_{CNT} \langle v_{CNT} \rangle d\omega}{ad} + \frac{1}{4}\int \omega N_{OX} f_{BE,OX} D_{OX} v_{ox} d\omega} \tag{3}$$

where $N$ refers to the atomic density (in atoms/cm$^3$ for SiO$_2$ and atoms/cm for CNTs), $v$ is the phonon velocity, $\omega$ is the phonon frequency, $f_{BE}$ is the Bose-Einstein (BE) distribution, and $D$ is the PDOS per atom as calculated by MD simulations.[16] We use the realistic PDOS rather than a Debye approximation because the latter has been previously found to cause large discrepancies with experimental data at high temperature.[36] In addition, the linear Debye approximation would not account for the quadratic CNT flexure modes.[37] The PDOS for a (10,10) nanotube with 1.37 nm diameter is calculated and shown in Fig. 3. Using the PDOS from CNTs of other diameters did not change our results significantly (presumably because the proportion of phonon modes remains approximately the same[16]), hence we used the PDOS shown in Fig. 3 as the phonon weighing function throughout the remainder of this work.

The phonon velocity in the amorphous SiO$_2$ is assumed to be isotropic and fitted with a single value,[38] $v_{ox}$ as shown in Table I. However, the CNT phonon velocity includes contributions from both the transverse and longitudinal polarizations along the c-axis (out-of-plane direction) of graphite, $v_c$.[34, 39] Not included are the a-axis (in-plane) modes which contribute minimally to thermal coupling in the geometry of interest, and are more relevant to vertical CNTs on a surface.[40] We note, however, that even for vertical CNTs some degree of tip bending must always exist, thus the geometry examined here and in Ref. 16 is likely to be most relevant. The value of $\langle v_{CNT} \rangle$ is derived from a geometrical averaging of $v_c$ over the shape of the CNT, described in Section V.B in more detail.

Knowing the transmission probability, we can now calculate the flux of phonons through the interface. This gives the thermal conductance per unit length from the CNT to SiO$_2$ as:

$$g = \frac{N_{CNT} \cdot b_t}{ad} \int \hbar\omega \frac{\partial f_{BE,CNT}}{\partial T} D_{CNT} \langle v_{CNT} \rangle \tau d\omega \tag{4}$$

where $b_t$ is the effective thermal contact width or footprint between the CNT and the substrate, to be determined by MD simulations (Fig. 1 and Fig. 5). This footprint is the effective width between CNT and the substrate over which heat is being transferred. Finally, to calculate a thermal boundary conductance that is comparable to experimental data, we must also include the effect of heat spreading into the oxide, given as:[41]

$$g_{ox} = \frac{2\pi \kappa_{ox}}{\ln\left(\frac{8t_{ox}}{\pi b_t}\right)} \qquad (5)$$

where $\kappa_{ox} \approx 1.4$ Wm$^{-1}$K$^{-1}$ is the SiO$_2$ thermal conductivity and $t_{ox} \approx 90$ nm is the SiO$_2$ layer thickness. This simple expression is appropriate when $t_{ox} \gg b_t$ as in our work, and the thermal spreading resistance contribution of the SiO$_2$ accounts for approximately 10-30% of the total thermal resistance. The total thermal conductance per unit length from CNT to ambient, as used in Eq. (2), is given by the simple thermal series network shown in Fig. 1:

$$g_{tot} = \left(\frac{1}{g} + \frac{1}{g_{ox}}\right)^{-1}. \qquad (6)$$

We note that any additional thermal spreading resistance into the Si wafer is negligible, and thus the Si wafer is assumed to be isothermal at $T_{Si} = 293$ K. Similarly, heat loss to ambient air can be neglected, where $g_{air} \sim 4\times10^{-4}$ WK$^{-1}$m$^{-1}$ has been previously estimated as an upper limit at one atmosphere,[42] three orders of magnitude lower than the heat loss to substrate.

## IV. Derivation of CNT Shape and Footprint

### A. Equilibrium Shape of a CNT

Nanotubes interact with the SiO$_2$ substrate through van der Waals (vdW) forces. In addition, our previous MD simulations[16] have shown that such CNTs do not remain rigid cylinders, but instead deform to minimize their overall vdW and curvature energy. Beyond a certain diameter CNTs relax to a compressed shape,[43, 44] which changes both their geometrical and equivalent thermal footprint on the substrate. To calculate the shape and thermal footprint for a wide variety of CNTs we employ MD simulations with a simplified Lennard-Jones (LJ) 6-12 potential:

$$V = 4\varepsilon\left[\left(\frac{\sigma}{r}\right)^{12} - \left(\frac{\sigma}{r}\right)^6\right]. \qquad (7)$$

Here, we simplify the SiO$_2$ substrate as a continuum plane. Therefore the collective vdW interaction per carbon atom situated at a height $h$ above an infinite half-space of SiO$_2$ can be approximated by the triple integral

$$V_{vdW} = \sum_{i=Si,O} \int_0^\infty \int_{-\infty}^\infty \int_{-\infty}^\infty 4\varepsilon_i n_i \left[\left(\frac{\sigma_i}{r}\right)^{12} - \left(\frac{\sigma_i}{r}\right)^6\right] dxdydz \qquad (8)$$

in which

$$r = \sqrt{x^2 + y^2 + (z+h)^2} \qquad (9)$$

and

$$\varepsilon_{Si} = 8.909 \text{ meV}, \sigma_{Si} = 3.326 \text{ Å}, n_{Si} = 0.0227 \text{ Å}^{-3}$$
$$\varepsilon_O = 3.442 \text{ meV}, \sigma_O = 3.001 \text{ Å}, n_O = 0.0454 \text{ Å}^{-3} . \qquad (10)$$

The values here are based on the Universal Force Field (UFF) model by Rappe et al[45] and were used in our previous MD simulations as well.[16] The integral in Eq. (8) can be evaluated analytically. It should be noted that this integral tends to give a lower bound estimate of the total interaction potential because it ignores the effects of local spikes of closely positioned atoms. The estimation error is reduced by assuming a relaxed configuration for the nearby silica molecules. Such an analysis gives

$$V_{vdW} = \sum_{i=Si,O} \frac{2\pi\varepsilon_i \sigma_i^3 n_i}{45} [2(\sigma_i/h)^9 - 15(\sigma_i/h)^3] \qquad (11)$$

which has a similar form as the original LJ potential, except with different exponents and prefactors. This effectively alludes to an $h^{-3}$ dependence of the vdW interaction potential. A plot of both the calculated potential and its second derivative (which is proportional to the interaction spring constant) is shown in Fig. 4.

For the covalent C-C interaction we used the empirical bond order Tersoff-Brenner potential.[46] In addition, we also used an intra-molecular LJ vdW potential with the following parameters for graphite[47] implemented via Eq. (7):

$$\varepsilon_C = 3.02 \text{ meV}, \sigma_C = 3.39 \text{ Å} . \qquad (12)$$

This potential for C-C interaction was found necessary to achieve proper equilibrium of the CNT shape. All MD simulations were carried out until the transient motions died off and a final steady-state solution was reached.

## B. Thermal Footprint of a CNT

To determine the thermal footprint of the CNT on $SiO_2$ we consider the square root of the second derivative of the vdW potential with respect to $h$ as heat transfer depends on this effective "spring constant" between the substrate and CNT. In particular, the thermal coupling is proportional to the vertical velocity component of the CNT atoms vibrating in the CNT-$SiO_2$ potential, itself proportional to the square root of this spring constant. Thus, to find the effective thermal footprint we used the square root of the CNT-$SiO_2$ spring constant to weigh the horizontal displacement $\Delta y$, as labeled in Fig 5(b), where the $y$-axis is centered at the middle of the CNT.

The thermal footprint ($b_t$) is different from the geometric footprint ($b_g$), the physical contact region between the CNT and substrate, both shown in Fig. 5. In the case of small diameter CNTs, the geometric footprint is of the order of the C-C bond length (see below), and the effective thermal footprint can be greater than the lateral width of the CNTs, i.e. their diameter. This occurs because in addition to the bottom half of the CNT conducting heat to the substrate, there is also thermal coupling from the top half of the CNT. The results of our simulations are shown in Fig. 5. Because MD simulations can be carried out for only one CNT of a particular diameter at a time, several were conducted for CNTs over a range of diameters 5–49 Å. We found the following quadratic function fit the simulation results of the thermal footprint for any diameter within the simulated range (Fig. 5a):

$$b_t = 0.037d^2 + 1.1d \qquad (13)$$

where both $b_t$ and $d$ are both in nanometers .

Our simulations also suggest that there are two different regimes represented by different equilibrium shapes of CNTs, as shown in Fig. 5. In the first regime ("I"), the diameter of the CNT is $d < 2.1$ nm and the curvature energy of the CNT is stronger than the vdW energy with the substrate. Thus in the first regime the cross-section of the nanotube is nearly circular, as shown on the left of Fig. 5(b) for a (7,7) CNT. In addition, the geometrical footprint (calculated by finding the furthest distance between the lowest points on the CNT) in this regime is extremely small and nearly constant at ~1.4 Å, the chemical bond length, as seen in Fig. 5(a).

In the second regime ("II") the diameter $d > 2.1$ nm, and the vdW energy with the substrate is stronger than the curvature energy of the CNT. Hence the final minimum energy shape of the CNT will be that of a deformed circle, as shown for a (22,22) CNT on the right of Fig. 5(b). In this regime the geometrical footprint begins to increase approximately linearly with diameter, as shown in Fig. 5(a). Another interesting observation is noted due to the repulsive nature of the vdW forces at very close distances, whose relative magnitudes are illustrated with arrows in Fig. 5(b). In this case, the bottom of the CNT is not perfectly flat. Instead the middle of the bottom region buckles up slightly, such that the force at the center is nearly zero. All these effects are captured in the thermal footprint calculation ($b_t$) fitted by Eq. (13) above, and used in the DMM thermal coupling simulations.

## V. Discussion

Figure 6(a) shows the directly measured power at breakdown ($P_{BD}$), and Fig. 6(b) displays the extracted TBC ($g$) vs. diameter $d$ for 29 metallic and semiconducting CNT devices where diameter was available. Fig 6(b) also includes modeling using the DMM described above (solid line) and the dashed lines fitted to MD simulations with vdW coupling strengths $\chi = 1$ and $\chi = 2$, as described in Ref. 16. Both data and modeling trends in Fig. 6(b) suggest that the TBC increases with diameter. The range of extracted $g$ corresponds to approximately the same order of magnitude previously extracted from thermal breakdowns.[11, 48] The representative set of vertical error bars on one m-CNT in Fig. 6(b) corresponds to a ±50 °C uncertainty in knowing the exact breakdown temperature. Horizontal error bars represent ±0.4 nm uncertainty in diameter from AFM measurements. Vertical error bars on the s-CNTs are derived as follows. The upper limit is set by assuming $L_{BD}/L = 0.75$ and the lower limit is set for the case of uniform heat generation (i.e. uniform field like m-CNTs). It is interesting to note that non-uniformity of heat generation plays a larger role in large diameter s-CNTs than in small diameter s-CNTs, perhaps due to the larger density of states in the former,[23] and the larger charge density variations it leads to.

### A. Dependence of Thermal Coupling on Diameter and Temperature

We observe that $g$ increases with diameter up to ~0.7 WK$^{-1}$m$^{-1}$ per unit length for the largest single-wall CNTs considered ($d \sim 4$ nm) at the breakdown temperature ($T_{BD} \sim 600$ °C). The diameter dependence of $g$ is primarily a result of the increase in thermal footprint, as shown in Fig 5(a). The thermal footprint is only expected to depend on CNT diameter, vdW coupling strength,[16] and SiO$_2$ surface roughness (see below), but not significantly on temperature since thermal expansion alone would lead to only ~0.1 % change in diameter[49] over the temperature range of interest. In Fig. 6(b) we also plot our previous MD simulations results.[16] The results from the MD simulations give lower values of thermal coupling $g$ because the DMM assumes, by definition, that all phonons are scattered diffusely at the interface[32] whereas this does not necessarily happen in MD simulations.

We also obtain the thermal contact conductance per unit area, $h = g/b_t$, as plotted in Fig. 6(c) and showing almost no dependence on diameter. From the breakdown experiments this value is in the range $h \approx$ 20–200 MWK$^{-1}$m$^{-2}$, whose upper range is comparable to that recently obtained[50] for graphene on SiO$_2$, i.e. ~100 MWK$^{-1}$m$^{-2}$. The DMM simulation predicts an upper limit for $h \approx$ 220 MWK$^{-1}$m$^{-2}$ with almost no diameter dependence. This appears to delineate the upper range of the $h$ values obtained experimentally.

We note that the empirically extracted and simulated TBCs in this study thus far are at an elevated temperature, given approximately by the CNT breakdown condition ($T_{BD}$ ~ 600 °C). To understand the effects of temperature on TBC, we plot our DMM model in Fig. 6(d) vs. temperature. This shows an expected increase in TBC with temperature, consistent both with graphene-SiO$_2$ experiments[50] and with CNT-SiO$_2$ MD simulations.[16] The thermal coupling per unit area *at room temperature* is thus ~130 MWK$^{-1}$m$^{-2}$, or approximately 40 percent lower than the thermal coupling near the CNT breakdown temperature. The upper limit of thermal coupling per unit length *at room temperature* is therefore ~0.4 WK$^{-1}$m$^{-1}$ for the largest diameters single-wall CNTs considered here ($d$ ~ 4 nm).

**B. Dependence of TBC on Phonon DOS and Velocity**

In addition to the thermal footprint, the PDOS of the SiO$_2$ as well as the distribution function ($f_{BE}$) also play a role in heat transport across the interface. We recall that the inset of Fig. 3 showed the calculated PDOS for both a (10,10) CNT and the SiO$_2$ substrate. While the nanotube contains a large PDOS peak at ~53 THz, this does not come into play directly because there are no equivalent high-frequency modes in the SiO$_2$. Fig. 3 also shows the Bose-Einstein distribution function at the CNT breakdown temperature ($T_{BD}$ ~ 600 °C). The distribution suggests very low occupation for all high frequency CNT modes. Since the Debye temperature for CNTs is very high, we expect that most substrates will serve as a low-pass filter for CNT phonons.

Aside from changing the thermal footprint, the deformed shape of the CNT also affects the average phonon velocity. This is a more subtle effect than that of diameter or surface roughness, but is included here for completeness. For instance, in the second regime ($d > 2.1$ nm) the CNT becomes flattened, leading to more atoms vibrating perpendicular to and closer to the substrate. Thus, the value $\langle v_{CNT} \rangle$ used in Eq. (3) represents this angle-averaged adjustment to $v_c$ listed in Table I, using the CNT shape suggested by MD simulations at each diameter (Fig. 5).

**C. Dependence of TBC on Surface Roughness**

There are several variables contributing to the spread of the experimental data shown in Figs. 6 and 7. The primary contributor is surface roughness. Since the value of $g$ is directly related to the contact area at the interface, an imperfect surface is roughly equivalent (pun intended!) to a decreased thermal contact area. Figure 7(a) replots the calculated TBC vs. diameter for a perfectly smooth surface (100%), for 75% of the maximum contact area, and for 50% of the maximum contact area. To analyze how surface roughness affects the spread directly, we experimentally find the root-mean-square (RMS) surface step height ($\Delta$) adjacent to the nanotube via AFM. However, intuitively we expect the *ratio* of diameter to roughness ($d/\Delta$) to be more important. In other words, we expect large diameter CNTs to be less affected by surface roughness than small diameter CNTs. Replotting $g$ as a function of $d/\Delta$ in Fig. 7(b), we observe that the scatter in the data becomes smaller than in Figs. 6(b) and 7(a). This supports our hypothesis and suggests the strong effect of surface roughness on thermal coupling at such nanoscale interfaces.

**D. Role of s-CNT vs. m-CNTs**

We note that the spread in m-CNTs breakdown data is smaller than in s-CNTs in Figs. 6 and 7. We believe this is due to threshold voltage ($V_{TH}$) shifting in s-CNTs during the high-field measurement process, which m-CNTs are essentially immune to. As the devices are swept to high $V_{SD}$ bias for breakdown, along with the applied gate bias ($V_{GD}$ = -15 V) this can lead to dynamic charge injection into the oxide, as studied in depth by Ref. 18. To understand the effect of threshold voltage on breakdowns, we plot the extracted $P_{BD}$ vs. initial $V_{TH}$ in Fig. 7(c), and find a slight but positive relationship. This suggests that in s-CNTs the variation in electronic behavior leads to the larger data spread, in addition to the variation due to surface roughness discussed above. Moreover, this also indicates a root cause which can render selective breakdown of m-CNTs (e.g. in CNT arrays[1,8] or networks[9]) as a challenging and imperfect approach. On one hand, the change in threshold voltage of s-CNTs due to dynamic charge injection into the $SiO_2$ at high field regions can turn them "on," allowing them to break down. On the other hand the variation in surface roughness itself cannot guarantee that all m-CNTs will break down at the same input power, or voltage.

**E. Comments on the Modeling Approach**

It is important to note that both the DMM and MD simulations employed in this work only capture the lattice vibration (phonon) contribution to thermal coupling. Nevertheless, the DMM in general appears to represent an upper limit to the spread of the experimental data which is otherwise lowered by effects like surface roughness. However, recent theoretical work has also suggested a possible electronic contribution to heat transport through coupling with surface phonon polaritons (SPPs) from the oxide.[51, 52] The SPP interaction drops off exponentially with the CNT-substrate distance, perhaps leading to a larger electronic contribution to heat transport in regime II of the CNT shape ($d > 2.1$ nm), where more CNT atoms are closer to the $SiO_2$ surface. However, since the SPP potential is strongly dependent on the interaction distance, it will also be affected by substrate surface roughness. Given these circumstances it is difficult to completely rule out energy relaxation through SPP scattering in practice, although we believe this appears to be significantly lower than the phonon coupling. In addition, any SPP contribution (however small) may be more notable in larger diameter CNTs ($d > 2.1$ nm) due to the increase in footprint.

Another mechanism for CNT-$SiO_2$ energy dissipation is inelastic phonon scattering at the interface, which is not captured by the DMM. Previously Chen et al[50] had compared an elastic DMM calculated by Duda et al[34] to the TBC between graphene and $SiO_2$ and found that the elastic DMM under-predicted the TBC by approximately an order magnitude. Hopkins[35] made a similar argument for inelastic scattering between acoustically mismatched materials. However our simulations do not differ from the data significantly, thus our calculations suggest that the contribution of inelastic scattering here is small.

**VI. Conclusions**

In summary, we have examined electrical breakdown and thermal dissipation between CNT devices and their $SiO_2$ substrate, the most common configuration found in CNT electronics. The breakdown location is found near the middle for m-CNTs and closer to the high field region at the drain for s-CNTs, consistent with the CNT temperature profile. In this context, thermal dissipation from CNT to $SiO_2$ dominates over dissipation at the CNT contacts. We found evidence of a direct relationship between the CNT-$SiO_2$ thermal boundary conductance (TBC) and the CNT diameter, in accord with previous MD simulations. To provide a more flexible means of analysis we developed a diffuse mismatch model (DMM) of the TBC using the full phonon density of states (PDOS). This approach appears to predict the upper limit of thermal transmission at the CNT-$SiO_2$ interface, and could be similarly applied to calculate the TBC of other dimensionally mismatched systems. Our experiments and modeling suggests a maximum TBC of ~0.4 $WK^{-1}m^{-1}$ per unit length at room temperature and ~0.7 $WK^{-1}m^{-1}$ at 600 °C for the largest diameter CNTs considered ($d$ = 3–4 nm). The maximum thermal conductance per unit area corresponds to approximately 130 $MWK^{-1}m^{-2}$ at room temperature and 220 $MWK^{-1}m^{-2}$ at 600 °C.

We have also studied the thermal footprint of a CNT through MD simulations which find the atomic configuration of lowest energy. These reveal two interaction regimes, the first one at smaller diameters ($d$ < 2.1 nm) where the CNT shape is dominated by its curvature energy, the other at larger diameters ($d$ > 2.1 nm) where the CNT shape is dominated by Van der Waals (vdW) coupling with the substrate. Finally, we found evidence that $SiO_2$ surface roughness strongly affects the TBC of such nanometer-sized interfaces. To improve CNT heat sinking applications, our results suggest the need to engineer ultra-flat surfaces, use large diameter CNTs, and find substrates with larger vdW coupling. To improve selective electrical breakdown of CNTs (e.g. metallic vs. semiconducting) it will also be essential to control the surface roughness of the substrate, as well as the threshold voltage of the semiconducting CNTs.

## VII. Acknowledgements

The authors would like to thank D. Cahill, J.-P. Leburton, V. Perebeinos, and J. Shiomi for valuable discussions. This work is supported by the Nanoelectronics Research Initiative (NRI), the NRI Hans Coufal Fellowship (A.L.), the NSF CAREER ECCS 0954423 award (E.P.), the MARCO MSD Focus Center (F.X.), and the NSF DMR 0504751 and CMMI 0906361 awards (R.A. and K.J.H.).

**References**

1. K. Ryu, A. Badmaev, C. Wang, A. Lin, N. Patil, L. Gomez, A. Kumar, S. Mitra, H. S. P. Wong, and C. W. Zhou, Nano Letters **9**, 189 (2009).
2. M. A. Panzer, G. Zhang, D. Mann, X. Hu, E. Pop, H. Dai, and K. E. Goodson, Journal of Heat Transfer **130**, 052401 (2008).
3. S. T. Huxtable, D. G. Cahill, S. Shenogin, L. Xue, R. Ozisik, P. Barone, M. Usrey, M. S. Strano, G. Siddons, M. Shim, and P. Keblinski, Nat Mater **2**, 731 (2003).
4. E. Pop, D. Mann, J. Cao, Q. Wang, K. Goodson, and H. J. Dai, Phys. Rev. Lett. **95**, 155505 (2005).
5. M. Lazzeri, S. Piscanec, F. Mauri, A. C. Ferrari, and J. Robertson, Phys. Rev. Lett. **95**, 236802 (2005).
6. M. A. Kuroda, A. Cangellaris, and J. P. Leburton, Phys. Rev. Lett. **95**, 266803 (2005).
7. P. C. Collins, M. S. Arnold, and P. Avouris, Science **292**, 706 (2001).
8. S. J. Kang, C. Kocabas, T. Ozel, M. Shim, N. Pimparkar, M. A. Alam, S. V. Rotkin, and J. A. Rogers, Nat Nano **2**, 230 (2007).
9. N. Patil, A. Lin, Z. Jie, W. Hai, K. Anderson, H. S. P. Wong, and S. Mitra, in *IEEE Intl. Electron Devices Mtg. (IEDM)* (IEEE, Baltimore, MD, 2009), p. 573.
10. E. Pop, Nanotechnology **19**, 295202 (2008).
11. E. Pop, D. A. Mann, K. E. Goodson, and H. J. Dai, J. Appl. Phys. **101**, 093710 (2007).
12. M. Steiner, M. Freitag, V. Perebeinos, J. C. Tsang, J. P. Small, M. Kinoshita, D. N. Yuan, J. Liu, and P. Avouris, Nature Nanotechnology **4**, 320 (2009).
13. J. Shiomi and S. Maruyama, Japanese Journal of Applied Physics **47**, 2005 (2008).
14. R. Prasher, Applied Physics Letters **90**, 143110 (2007).
15. H. Y. Chiu, V. V. Deshpande, H. W. C. Postma, C. N. Lau, C. Miko, L. Forro, and M. Bockrath, Physical Review Letters **95**, 226101 (2005).
16. Z.-Y. Ong and E. Pop, Phys. Rev. B **81**, 155408 (2010).
17. A. Liao, Y. Zhao, and E. Pop, Physical Review Letters **101**, 256804 (2008).
18. D. Estrada, A. Liao, S. Dutta, and E. Pop, Nanotechnology **21**, 085702 (2010).
19. J. Lee, A. Liao, E. Pop, and W. P. King, Nano Letters **9**, 1356 (2009).
20. Z. Yao, C. L. Kane, and C. Dekker, Physical Review Letters **84**, 2941 (2000).

[21] P. G. Collins, M. Hersam, M. Arnold, R. Martel, and P. Avouris, Phys. Rev. Lett. **86**, 3128 (2001).

[22] Y. F. Chen and M. S. Fuhrer, Physical Review Letters **95**, 236803 (2005).

[23] Y. Zhao, A. Liao, and E. Pop, IEEE Electron Device Letters **30**, 1078 (2009).

[24] K. Hata, D. N. Futaba, K. Mizuno, T. Namai, M. Yumura, and S. Iijima, Science **306**, 1362 (2004).

[25] N. Y. Huang, J. C. She, J. Chen, S. Z. Deng, N. S. Xu, H. Bishop, S. E. Huq, L. Wang, D. Y. Zhong, E. G. Wang, and D. M. Chen, Physical Review Letters **93**, 075501 (2004).

[26] E. Pop, D. Mann, Q. Wang, K. Goodson, and H. J. Dai, Nano Letters **6**, 96 (2006).

[27] I. K. Hsu, R. Kumar, A. Bushmaker, S. B. Cronin, M. T. Pettes, L. Shi, T. Brintlinger, M. S. Fuhrer, and J. Cumings, Applied Physics Letters **92**, 063119 (2008).

[28] V. V. Deshpande, S. Hsieh, A. W. Bushmaker, M. Bockrath, and S. B. Cronin, Physical Review Letters **102**, 105501 (2009).

[29] Y. Ouyang and J. Guo, Applied Physics Letters **89**, 183122 (2006).

[30] F. Xiong, A. Liao, and E. Pop, Applied Physics Letters **95**, 243103 (2009).

[31] L. Shi, J. H. Zhou, P. Kim, A. Bachtold, A. Majumdar, and P. L. McEuen, J. Appl. Phys. **105**, 104306 (2009).

[32] E. T. Swartz and R. O. Pohl, Reviews of Modern Physics **61**, 605 (1989).

[33] R. Prasher, Physical Review B **77**, 075424 (2008).

[34] J. C. Duda, J. L. Smoyer, P. M. Norris, and P. E. Hopkins, Appl. Phys. Lett. **95**, 031912 (2009).

[35] P. E. Hopkins, Journal of Applied Physics **106**, 013528 (2009).

[36] P. Reddy, K. Castelino, and A. Majumdar, Applied Physics Letters **87**, 211908 (2005).

[37] G. D. Mahan and G. S. Jeon, Physical Review B **70**, 075405 (2004).

[38] P. G. Sverdrup, Y. S. Ju, and K. E. Goodson, Journal of Heat Transfer **123**, 130 (2001).

[39] K. Sun, M. A. Stroscio, and M. Dutta, Superlattices and Microstructures **45**, 60 (2009).

[40] R. Prasher, T. Tong, and A. Majumdar, Journal of Applied Physics **102**, 104312 (2007).

[41] F. P. Incropera and D. P. DeWitt, *Fundamentals of Heat and Mass Transfer* (Wiley, New York, 2001).

[42] D. Mann, E. Pop, J. Cao, Q. Wang, and K. Goodson, Journal of Physical Chemistry B **110**, 1502 (2006).

[43] T. Hertel, R. E. Walkup, and P. Avouris, Physical Review B **58**, 13870 (1998).

[44] S. Zhang, R. Khare, T. Belytschko, K. J. Hsia, S. L. Mielke, and G. C. Schatz, Physical Review B **73**, 075423 (2006).

[45] A. K. Rappe, C. J. Casewit, K. S. Colwell, W. A. Goddard, and W. M. Skiff, Journal of the American Chemical Society **114**, 10024 (1992).

[46] D. W. Brenner, Physical Review B **42**, 9458 (1990).

[47] B. Chen, M. Gao, J. M. Zuo, S. Qu, B. Liu, and Y. Huang, Appl. Phys. Lett. **83**, 3570 (2003).

[48] H. Maune, H. Y. Chiu, and M. Bockrath, Applied Physics Letters **89**, 013109 (2006).

[49] P. K. Schelling and R. Keblinski, Physical Review B **68**, 035425 (2003).

[50] Z. Chen, W. Jang, W. Bao, C. N. Lau, and C. Dames, Appl. Phys. Lett. **95**, 161910 (2009).

[51] A. G. Petrov and S. V. Rotkin, JETP Letters **84**, 156 (2006).

[52] S. V. Rotkin, V. Perebeinos, A. G. Petrov, and P. Avouris, Nano Letters **9**, 1850 (2009).

**Figures and Tables**

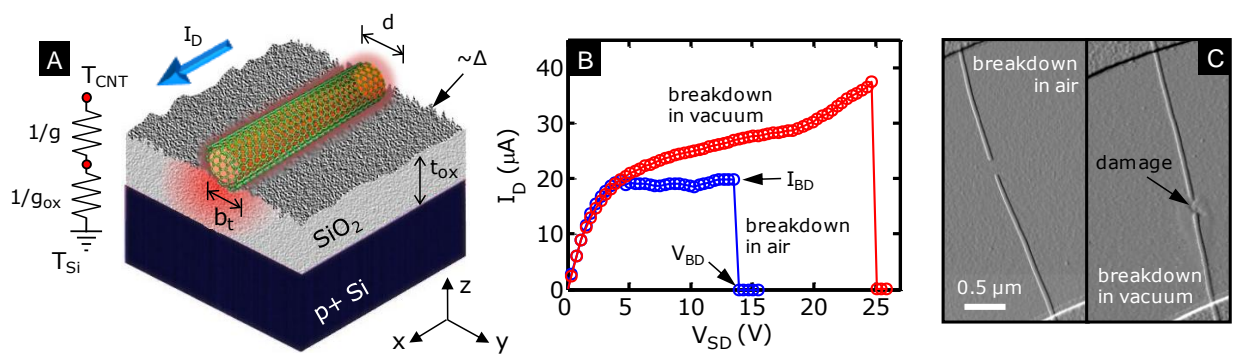

**Figure 1:** (Color online) (a) Schematic cross-section of typical CNT device with diameter $d$ and thermal footprint $b_t$ (also see Fig. 5) on SiO$_2$ substrate with thickness $t_{ox}$ and surface roughness $\Delta$. The p+ silicon is used as a back-gate. The device layout with source and drain terminals is shown in Fig. 2(a). As current ($I_D$) passes in the CNT, the generated Joule heat dissipates through the substrate. The equivalent thermal circuit includes CNT-SiO$_2$ interface thermal resistance ($1/g$) and spreading resistance in the SiO$_2$ ($1/g_{ox}$). (b) Typical electrical breakdown of similar CNTs shows higher breakdown power in vacuum (~$10^{-5}$ torr) than in ambient air. This illustrates the role of oxygen for CNT breakdown in air. (c) Atomic force microscopy (AFM) images of CNTs broken in air (left) and vacuum (right). Due to higher power, breakdowns in vacuum can lead to SiO$_2$ surface damage, which is not observed for air breakdowns.

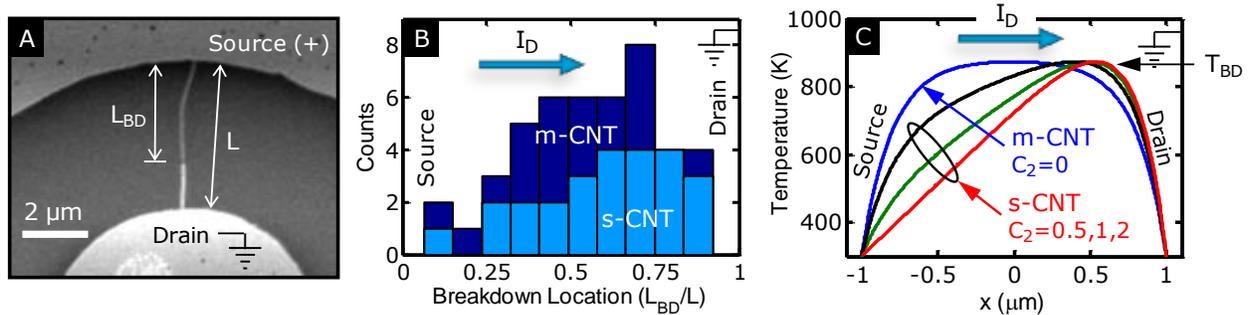

**Figure 2:** (Color online) (a) Scanning electron microscope (SEM) image of CNT device showing breakdown location ($L_{BD}$). (b) Histogram of breakdown location normalized by CNT length ($L_{BD}/L$) indicating the majority of m-CNTs break near the middle, and s-CNTs break closer to the grounded drain. (c) Computed temperature distribution at breakdown (maximum temperature = $T_{BD}$) along a 2 μm long CNT with Eq. (2) using $C_1 = 1$ and varying $C_2$. $C_2 = 0$ corresponds to m-CNTs (uniform heat dissipation) and $C_2 > 0$ corresponds to s-CNTs. For s-CNTs biased under hole conduction the heat generation and temperature profile are skewed towards the grounded drain terminal. Block arrows in (b) and (c) show direction of hole flow.

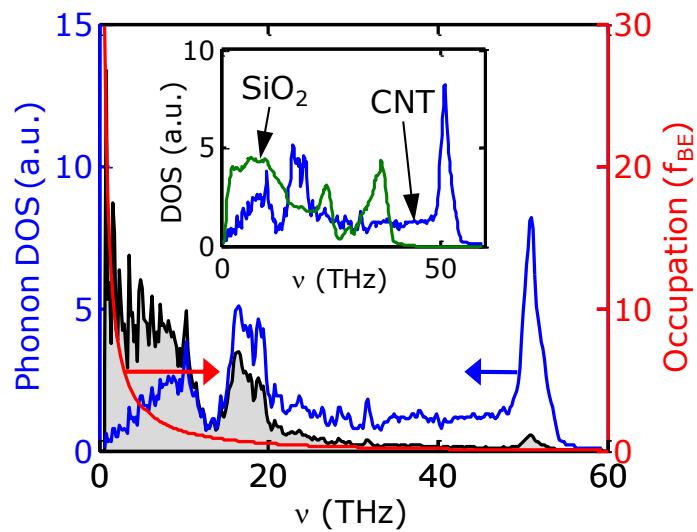

**Figure 3:** (Color online) The phonon density of states (PDOS) for a (10,10) nanotube from MD simulations.[16] The Bose-Einstein occupation ($f_{BE}$) at room temperature is plotted in red against the right axis. Shaded in gray is the product of the PDOS with $f_{BE}$, showing diminished contribution from higher frequency phonon modes. The inset shows the PDOS of the CNT and that of $SiO_2$, the latter displaying a lower cutoff near 40 THz.

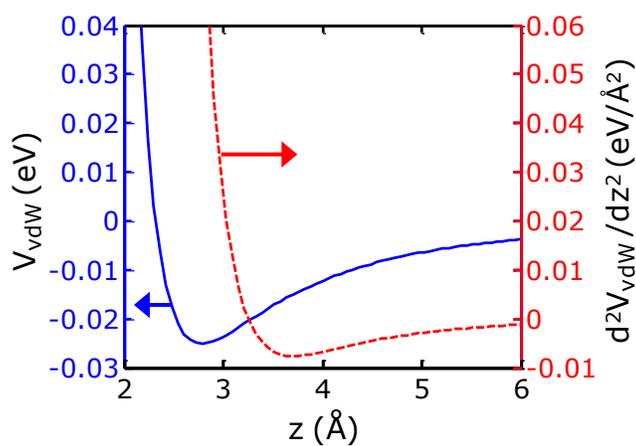

**Figure 4:** (Color online) Van der Waals potential (blue solid line) interaction between CNT and $SiO_2$, as used in calculations to derive the thermal footprint (Fig. 5). The second derivative of the potential (red dashed line) with respect to distance from the surface ($z$) represents a spring constant of the interatomic interaction. The square root of this spring constant is used to weigh the contribution of each atom to the effective thermal footprint ($b_t$) of the CNT.

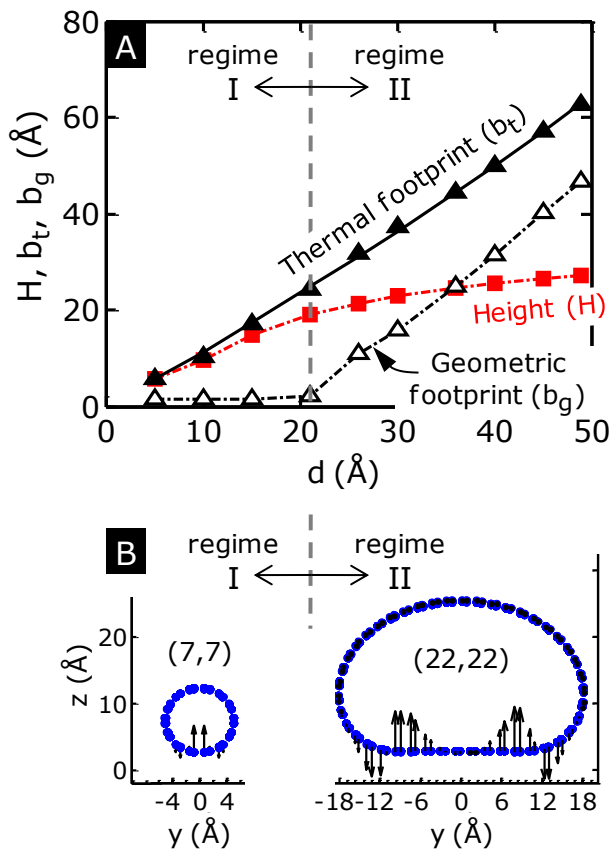

**Figure 5:** (Color online) (a) Nanotube height $H$ (■), geometric footprint $b_g$ (Δ), and thermal footprint $b_t$ (▲) on the SiO$_2$ substrate as a function of CNT diameter, from MD simulations. A fit to the thermal footprint is shown as a solid line from Eq. (13). (b) Calculations reveal two distinct regimes: in regime I ($d < 2.1$ nm) the CNT shape is nearly circular, dominated by the curvature energy; in regime II ($d > 2.1$ nm) the CNT shape becomes flattened, with a stronger influence of the surface vdW interaction. Example CNTs from each regime are shown, (7,7) with $d = 9.6$ Å for regime I, and (22,22) with $d = 30.6$ Å for regime II. Small vertical arrows indicate the relative magnitude of vdW force coupling with the substrate at each atomic position. The SiO$_2$ surface is at $z = 0$. The minimum CNT-SiO$_2$ distance is 2.5 Å at the middle of the (7,7) CNT and toward the sides of the (22,22) CNT. The middle of the (22,22) CNT is slightly "buckled" with 2.8 Å separation from SiO$_2$ (also noted by the lower vdW force there).

...

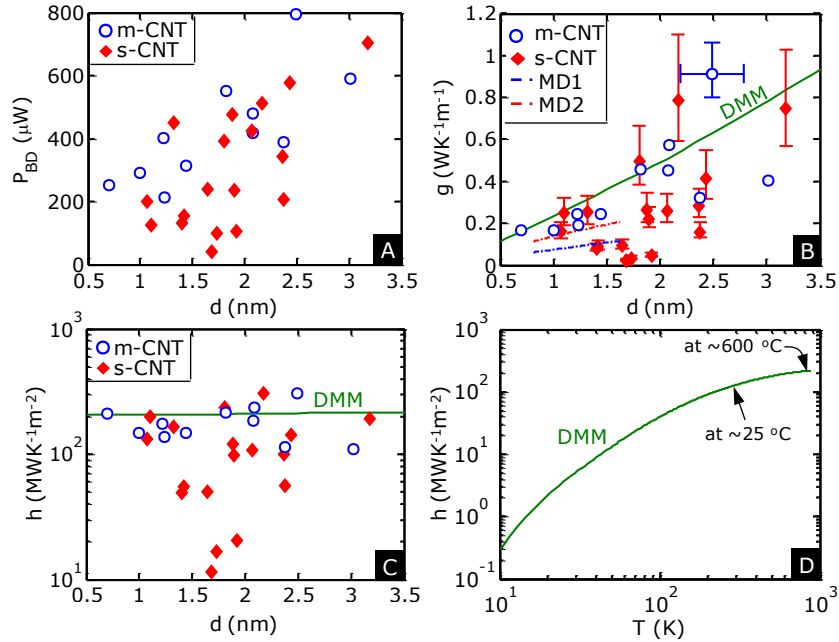

**Figure 6:** (Color online) (a) Electrical breakdown power (in air) of CNTs vs. diameter $d$, showing proportional scaling. (b) Extracted CNT-$SiO_2$ thermal coupling $g$ vs. $d$ (see text) for both metallic (m) and semiconducting (s) CNTs. Solid line is the DMM calculation and dash-dotted lines are fitted to MD simulations with different vdW coupling strengths ($\chi=1$ and $\chi=2$ respectively, from Ref. 16). (c) CNT-$SiO_2$ thermal coupling per unit area $h$ vs. $d$, showing the DMM represents an upper-limit scenario of heat dissipation. The spread in the data and lower apparent thermal coupling in practice is attributed to $SiO_2$ surface roughness, and charge trapping near semiconducting CNTs (see text). (d) Calculated temperature dependence of the upper limit thermal coupling per unit area. Thermal coupling at room temperature (~130 $MWK^{-1}m^{-2}$) is ~40% lower than at the breakdown temperature (~220 $MWK^{-1}m^{-2}$).

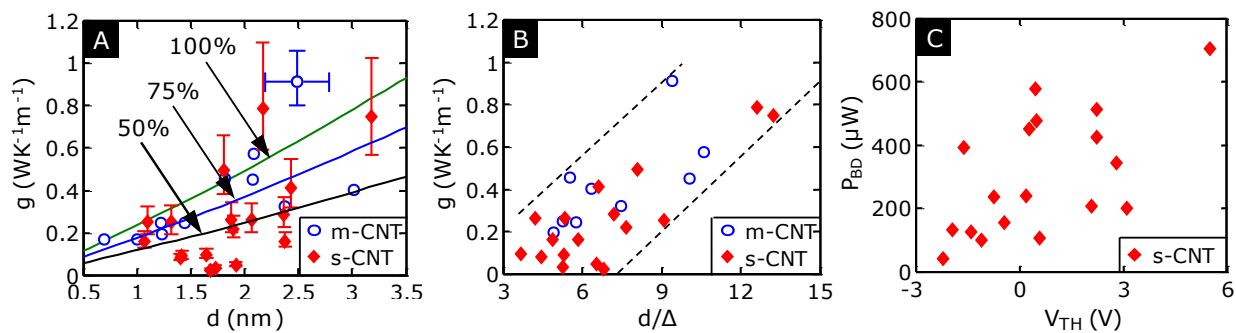

**Figure 7:** (Color online) (a) CNT-SiO$_2$ thermal coupling $g$ vs. diameter $d$ (symbols = data) and DMM simulations (lines) for perfect substrate contact (100%), and for 75% and 50% effective contact area due to SiO$_2$ surface roughness (also see Fig. 1). (b) Replot of same experimental data vs. diameter scaled by RMS surface roughness ($d/\Delta$) measured by AFM near each CNT. This indicates the role of SiO$_2$ surface roughness for thermal dissipation from CNTs. Dashed lines are added to guide the eye. (c) Breakdown power $P_{BD}$ for semiconducting CNTs (s-CNTs) alone plotted with respect to their threshold voltage ($V_{TH}$). The variance in $V_{TH}$ is also a contributing factor to the spread in extracted thermal coupling data for s-CNTs.

| Parameter | Value |
|---|---|
| $v_c$ | 932 m/s |
| $v_{ox}$ | 4.1 km/s |
| $N_{CNT}$ | 16.3 atoms/Å |
| $N_{ox}$ | 0.0227 molecules/Å$^3$ |
| $T_{BD}$ | 873 K |

**Table I:** Parameters used in the DMM model.